\documentstyle[aps,preprint,eqsecnum]{revtex}
\def\prl{Phys. Rev. Lett. }

\def\be{\begin{equation}}
\def\ee{\end{equation}}

\begin{document}

\title{Avoided Critical Behavior in a Uniformly Frustrated System}

\author{L.~Chayes}
\address{
Department of Mathematics
UCLA
Los Angeles, CA 90095, USA}

\author{V.~J. ~Emery}
\address{
Deptartment of Physics,
Brookhaven National Laboratory,
Upton, NY  11973, USA}

\author{S.~A.~Kivelson, Z.~Nussinov }
\address{
Deptartment of Physics,
UCLA,
Los Angeles, CA  90095, USA}

\author{G.~Tarjus}
\address{  Laboratoire de
Physique Th{\'e}orique des Liquides,
Universit{e} Pierre et Marie Curie,
75252, Paris Cedex 05, France}

\maketitle
\begin{abstract}
We study the effects of
weak long-ranged antiferromagnetic interactions of strength $Q$
on a spin
model with predominant short-ranged ferromagnetic interactions.
In three dimensions, this
model exhibits an avoided critical point in the sense that
the critical temperature $T_c(Q=0)$
is strictly greater than $\lim_{Q\to 0} T_c(Q)$.  The
behavior of this system at
temperatures less than $T_c(Q=0)$ is controlled by the
proximity to the avoided critical point.  We also quantize the
model in a novel way to study the interplay between charge-density
wave and superconducting order.

\end{abstract}
\narrowtext

\section{Introduction}

There are many diverse contexts in which a short-ranged tendency
to order is thwarted by a long-ranged frustrating interaction.  In
particular, recent theories of the glass transition\cite{glass}
and of the
properties of doped Mott insulators\cite{EKL,EK}
have lead to the consideration of
such models.  In the first example, a tendency of the molecules in a
supercooled liquid to pack into a locally preferred structure is
frustrated by the inability of such structures to tile
space;  the long-range nature of the induced interaction represents
the superextensive growth of strain which would occur if the locally
preferred structure were forced to tile space.  In the second
example, a short-range tendency of holes in an antiferromagnet to
phase separation competes with the long-range Coulomb repulsion
between holes.  In this latter case, effects of quenched disorder and
quantum fluctuations may also be important.

It has been argued\cite{glass}
that the effect of such uniform, long-range
frustration is an avoided critical point, leading to a phase diagram
of the sort shown in Figure 1.  Here, $T$ is the temperature and
$Q$ is the strength of the frustrating interaction. The salient feature
of this phase diagram is the discontinuity in $T_c$ in the
$Q\rightarrow 0$ limit.  The presence of the avoided critical point
leads to new types of fluctuation phenomena for $Q \ll 1$
and $T\lesssim T_c(Q=0)$.

To capture the essential physics of these problems, we will consider
a spin model on the $d$-dimensional hypercubic lattice.  Our starting
point is the classical Hamiltonian
\be
H_{cl}[S]= \frac J 2
\sum_{<\vec R,\vec R'>} \left|S_{\vec R}- S_{\vec R'}\right|^2 +
\frac Q 2 \sum_{\vec R\ne \vec R'} v(\vec R-\vec R')
S_{\vec R} \cdot S_{\vec R'}
\ee
where $J$ and $Q$ are positive couplings, the sum over $<\vec R,\vec R'
>$ is over
nearest neighbor pairs, and $v(\vec R)$ is long-ranged, with asymptotic
falloff
\be
v(\vec R) \sim | \vec R|^{-x}
\ee
with $0 < x \leq d$.  Specifically, it is most convenient to
express the interaction on the lattice in terms of its Fourier transform,
which in the simplest cases we will consider here, implies that the
Hamiltonian is
\be
H_{cl}[S]=\frac 1 2
\sum_{\vec k} {\cal J}(\vec k) \left|\tilde S_{\vec k}\right|^2.
\label{eq:Hcl}
\ee
Here $\tilde S$ is the Fourier transform of $S$,
\be
{\cal J}(\vec k) = J \Delta(\vec k) + Q  \left[\Delta(\vec k)\right]^{-y}
+ ...\ \
\label{eq:calJ}
\ee
with $y=(d-x)/2$, and
\be
\Delta(\vec k) = \sum_{a=1}^d V(k_a)
\ee
is the Fourier transform
of the lattice Laplacian, and
\be
V(k) = 2\left[1-\cos(k)\right]\approx k^2 \ \ {\rm for} \ \ k \ll 1.
\ee
In Eq. (\ref{eq:calJ}), the
$...$ refers to additional, short-range terms which
are generically present and which
will be included for reasons that will become clear shortly.
To make our discussion explicit, and because it is typically the case
of most physical interest,
we will focus on the Coulombic case $y=1$;
the principal results are qualitatively
similar for any long-range potential.

This model can be studied with
various definitions of the spin variables, $S_{\vec R}$.
Probably, the most interesting case
is
the Ising version,\cite{ute,chandler,etc} where $S_{\vec R}=\pm 1$.
The majority of the results in the present paper concern the exactly
solvable
``mean spherical'' version of this model, in which the spins are taken to
be real numbers with the mean global constraint
\be
N^{-1}\sum_{\vec R} < \left[S_{\vec R}\right]^2> = 1
\label{eq:constraint}
\ee
where $N$ is the number of sites.  As is well known,
this model is equivalent,
in the thermodynamic limit, to the usual spherical model in which
this global constraint is enforced configuration  by configuration
\cite{lw}  and to the large
$n$ limit of the $O(n)$ model.  \cite{largen}
(Ongoing work on the $1/n$ expansion will shed some additional light on the
relation of the present results to the properties of the model at finite $n$.
\cite{zohar})

Since, in the context of doped Mott insulators,
quantum effects are important, we also consider a quantum version
of this model.  In particular, we wish to consider a model with
two distinct types of low
temperature ordered phases:
those with spin (charge) order {\it and}
those with momentum (superconducting) order.
These order parameters are
canonically conjugate and dual to each other.
We are motivated in this choice by an analogy with
models of hard-core bosons on a lattice, or equivalently to a
spin $s=1/2$ frustrated quantum Heisenberg-Ising model in which
the $XY$ coupling is the particle kinetic energy (and
hence associated with the momentum) while
the Ising ordering is associated with ordering of the dual fields,
(charge ordering)
which can either correspond to phase separation or formation
of a charge density wave ordered state.\cite{1dexample}
With this in mind,  we define a quantum model
\be
H=H_{cl}[S] + H_{qu}[P],
\label{eq:quantum}
\ee
where $H_{qu}$ and $H_{cl}$ are appropriate quadratic forms,
in which the ``momenta'', $\{P_{\vec R}\}$, are cannonically
conjugate to the spins, $\{S_{\vec R}\}$,  {\it i.e.}
\be
[S_{\vec R},P_{\vec R'}]=i\hbar \delta_{\vec R,\vec R'}.
\ee
The constraint equation
is
\be
N^{-1}\sum_{\vec R} \left[ \alpha_s\left< |S_{\vec R}|^2\right>+
\alpha_p\left< |P_{\vec R}|^2\right> \right] =s^2.
\label{eq:quconstraint}
\ee
So long as neither $\alpha_s$ nor $\alpha_p$ is zero, we can,
without loss of generality,\cite{alpha} take $\alpha_s=\alpha_p=1$.
To be concrete, we also
confine our considerations to the simplest case in which
$H_{qu}$ consists of the simplest, unfrustrated nearest-neighbor
ferromagnetic interaction
\be
H_{qu}[P]=\frac W 2\sum_{<\vec R,\vec R'>}
\left|P_{\vec R}- P_{\vec R'}\right|^2.
\label{eq:hqu}
\ee
Thus, $W$ is the bare superfluid stiffness.
It also is necessary to augment the Hamiltonian with a uniform
field, $K$, which favors charge ordering ($<S_{\vec R}> \ne 0$)
when negative and superconducting ordering
(($<P_{\vec R}> \ne 0$) when positive,
\be
H_{cl} \rightarrow H_{cl} +\frac K 2 \sum_{\vec R} |S_{\vec R}|^2,
\label{eq:defK}
\ee
or, in other words, ${\cal J}(\vec k) \to {\cal J}(\vec k) + K$.

In the context of quantum spin glasses,
Nieuwenhuizen
\cite{dutch} and Hartman and Weichman\cite{pw}
have considered quantized versions of the spherical model.
In both cases, after some manipulation,\cite{dutch,pw} it is possible
to write the Hamiltonian in the form of Eqs. (\ref{eq:quantum}
- \ref{eq:hqu}).
However, the constraint considered by Hartman and Weichman corresponds
to the case $\alpha_s=0$, while that of Nieuwenhuizen corresponds
to $\alpha_p=0$. As we shall see, this difference has important
physical consequences.  In particular, to exhibit the two
conjugate ordered phases,
it is necessary that both $\alpha$'s be non-zero.

Effects of quenched
disorder can be included by adding a random field to the
Hamiltonian
\be
H_{dis}=\sum_j h_j S_j.
\ee
For our purposes, the ensemble of random fields is adequately specified by
its second moment,
\be
{\bf [}\tilde h_{\vec k} \tilde h_{\vec k'} {\bf ]}
=\delta_{\vec k,-\vec k'} f(\vec k),
\ee
where the ${\bf [ \ \ ] }$ signifies configuration averaging.
(We imagine that ${\bf [} h_j {\bf ]}=0$, although this is unimportant.)
Since the random fields are, presumably, generated by remote quenched charges,
we will typically suppose that $f(\vec k)$ vanishes as $k \rightarrow 0$ as
$f(\vec k) \sim k^2$, but is otherwise positive at all other values
of $\vec k$.
(Note that the two previously cited studies\cite{dutch,pw} considered
quantum spherical models with
random exchange
interactions so as to construct a spherical version
of a spin-glass;  this problem is interesting but considerably more
complex than the one considered here.)

We now return to the additional terms (signified by $...$) in Eq.
(\ref{eq:Hcl}).
The Hamiltonian, as written, depends on $\vec k$ only through
$\Delta(\vec k)$.  It turns out that this implies a degeneracy
(that is non-generic
for lattice systems) which
is a consequence of
a lattice version of the rotational symmetry
of free space and hence
would be exact in a continuum version of the model.
We will lift this degeneracy through the inclusion
of the term
\be
...\ \ = \frac 1 2 \lambda\left\{
\sum_{a\ne b}V(k_a)V(k_b)\right\}
\label{eq:lambda}
\ee
which in position space corresponds to a second neighbor interaction.
This term vanishes like $k^4$ as $k \to 0$;
notice that, to this order, any perturbation consistent with the
symmetry of the hypercubic lattice can be written as $\lambda^{\prime}
\left[\Delta(\vec k)\right]$ plus the term in Eq. (\ref{eq:lambda}).
However, any portion of the perturbation which depends only on the
Laplacian does not lift the degeneracy.  In this sense,
Eq. (\ref{eq:lambda}) is the unique, leading
order term
that breaks the ``rotational'' symmetry of the continuum.
For concreteness, we will always consider the
model with $\lambda$ positive, unless otherwise stated.

To summarize, in the rest of the paper we will consider the statistical
mechanics of the classical model defined by the Hamiltonian in Eq.
(\ref{eq:Hcl}) with
\be
{\cal J}(\vec k) = J \Delta(\vec k) + Q  \left[\Delta(\vec k)\right]^{-y}
+ \frac 1 2 \lambda \sum_{a\ne b}V(k_a)V(k_b) + K
\label{eq:calJtot}
\ee
and with the equation of constraint defined in Eq. (\ref{eq:constraint}).  We
will then consider the quantized version of the same model defined by
the Hamiltonian in Eq. (\ref{eq:quantum}) with $H_{qu}$ defined by
Eq. (\ref{eq:hqu}) and the equation of constraint by Eq.
(\ref{eq:quconstraint})
with $\alpha_s=\alpha_p=1$.
Henceforth,
we will work in units such that  Planck's constant,
Boltzmann's constant, the energy scale $J$,
and the lattice constant are all set equal to 1.

\section{Discussion of Results}

\subsection{Classical Model}

For $Q=0$, the model described is the usual spherical model.  As is well
known for this
case, $d=2$ is the lower critical dimension;  there is no finite temperature
transition in $d=2$ but the correlation length diverges rapidly, like
$\xi_0=L(\lambda)\exp\left[{2\pi /T}\right]$,
as $T\rightarrow 0$, which is qualitatively
similar to the behavior of the $O(n)$ model for $n >2$.  In $d=3$,
there is a finite temperature transition at a critical temperature
$T_c=A(\lambda)$ where $A(\lambda)$ is a strongly varying function of
$\lambda$ with $A(0)\simeq 0.79$, and the correlation length diverges as
$\xi_0=(4\pi/A)
\left[(T-T_c)/T_c\right]^{-\nu}$ with $\nu=1$ as the temperature
approaches $T_c$ from above.  The spherical model is somewhat
unsatisfactory for temperatures below $T_c$ where there
is a well
defined growth of the order parameter,
$\left<S_j\right> = \left[(T_c-T)/T_c\right]^{1/2}$, but
the (Josephson) correlation length is infinite.
(This is one of the problems we hope will be fixed when $1/n$ corrections
are included.\cite{zohar})

For non-zero $Q$, ferromagnetic order is completely forbidden.  However,
modulated order of one sort or another is permitted.  For small $Q$, it can
be easily seen by minimizing
${\cal J}(\vec k)$ in Eq. (\ref{eq:Hcl}) with respect
to $k$ that the preferred order occurs for $k=q$, where $V(q)
= \sqrt{Q}$.  Since
$q$ is small in this limit, a first approximation to the physics can
be obtained by taking a continuum limit, in which only the order $k^{-2}$ and
$k^2$
terms  in ${\cal J}$ are retained;  in particular, this means
that the term proportional to
$\lambda$ is neglected since it is
smaller by a factor of $\sqrt{Q}$ than the first two
terms for $k \sim q$.  However, as is well known in the
continuum theory of smectic liquid crystals, the transverse fluctuations of
such density wave order are sufficient to destroy the long-range order which,
in the context of the spherical model is equivalent to the destruction
of any finite temperature transition.  (A power law phase, as in the
2$d$ XY model or the 3$d$ continuum theory of smectic liquid crystals is
impossible for this simple class of models.)  In the present
context, this is a consequence of the existence of a co-dimension
one hypersurface of minimizing wave vectors which implies insufficient
stiffness against
fluctuations to permit a finite temperature ordering transition in any
dimension.  As discussed above, this piece of continuum physics is
reproduced on the lattice when only the first two terms in ${\cal J}$ are
retained (even though they are computed to all orders in $k$). In this
case the preferred order occurs at $\vec k$ on the hypersurface $\Delta(\vec k)
=\sqrt{Q}$ and therefore so long as $0 < \sqrt{Q} < 4d$, no finite temperature
ordering transition is possible.
(For larger $Q$, even if $\lambda=0$, the lattice asserts itself,
and ordering can occur.)
Thus we see, trivially, that in the
model with $\lambda =0$, there is an avoided critical point for all
dimensions $d > 2$, as in Fig. 1, in which $T_c$ drops from a finite value
for $Q=0$, to zero for $Q\ne 0$.

We now address the question of what happens to this phase diagram when
the symmetry-breaking
term proportional to $\lambda$ is introduced.
In this case, the minima of ${\cal J}(\vec k)$ occur
at isolated points in the Brillouin zone, and hence finite-temperature
ordering is possible at non-zero $Q$ in all dimensions $d > 2$.  For
$Q<16$ and $\lambda$ positive, there are $2d$
minima which occur at $\vec k = \pm q \hat e_a$,
for $a=1, ..., d$, with $V(q)=\sqrt{Q}$,
and hence correspond to unidirectional striped
phases.
(The ordering vectors for all $Q$ and all $\lambda$ are
listed in Table 1.)
In dimension $d=3$, as shown in Fig. 1, there remains a finite discontinuity
in $T_c$ as $Q\rightarrow 0$.  Specifically, for positive $\lambda$,
$T_c(Q=0) = A_o + A_1\lambda
+{\cal O}(\lambda^2)$ with $A_o=A(0)$ and
$A_1=A^{\prime}(0)\simeq 0.074$, while
\be
\lim_{Q \rightarrow 0}
T_c(Q) = {\sqrt{\lambda} T_c(0) \over
BT_c(0) +\sqrt{\lambda}}
\ee
with $B\simeq 0.14245$.
For $d > 3$,
finite $\lambda$ results in a continuous, although extremely non-analytic
behavior of $T_c$ at small $Q$:  $T_c(Q)=T_c(0)[1 - B_d T_c(0)
\lambda^{-\frac 1 2}(Q)^{(d-3)/4} ]
$ where $B_d$ is a dimension dependent number of order 1.
(For what it is worth, for $2<d<3$, there is a finite discontinuity in
$T_c$, and indeed, $T_c$ is finite for $Q=0$, but tends to zero in
the $Q \rightarrow 0$ limit as $T_c \sim B_d^{-1} \sqrt{\lambda} (Q)^{(3-d)/4}
$.)

In Fig. 1, we show the phase diagram for the three-dimensional classical
model as a fuction of $Q$ and
$T$ for fixed, positive $\lambda$ (The solid lines in the
figure are the phase boundaries computed for
$\lambda = 1/4$, but the  qualitative results are insensitive to the value
of $\lambda$.)  There appear two dashed lines in the small-$Q$ region of the
figure which signify crossover temperatures discussed below:

\noindent{$\bullet$}  $T_1$  marks the
temperature at which the frustration becomes significant.  At
$T > T_1$, correlations behave essentially like those of the
model with $Q=0$, while for $T < T_1$, the effect of the
frustration is to break the incipient ferromagnetic order into randomly
oriented ``domains'' of typical size, $\xi$, which at $T\approx T_c(Q=0)$
has magnitude
$\xi\approx (1/Q)^{1/4}$, and which grows
slowly as the temperature is lowered.

\noindent{$\bullet$}  $T_2$ marks the temperature at which lattice effects
becomes
important.   For $T > T_2$, the fluctuations are essentially
isotropic, while for $T < T_2$, the  correlation length begins to diverge as
the
ordering temperature is approached, with the same critical exponent as the
unfrustrated model, and the correlation functions begin to choose  preferred
orientations for the domains, corresponding to the onset of stripe ordering.
One additional pathology of the spherical model apparent in this phase diagram
(which, we expect, is corrected in order $1/n$) is that there is no
commensurate
lock-in whatsoever for small $Q$, {\it i.e.} there is no preference for stripe
ordering wave vectors $q$ which are commensurate
with the underlying lattice.
In Fig. 3 we show the Fourier transform,
$\tilde G(\vec k)$, of the spin-spin correlation function,
\be
G(\vec R) = \left < S_{\vec 0} S_{\vec R}\right>
\ee
at various temperatures for a fixed, small value of $Q=1/2$ and
$\lambda=1/4$.  One can
clearly see
the differences in the structure of the correlations in the different
regimes of temperature.

The most dramatic manifestation of avoided critical behavior is the existence
of these sharp crossover regimes.  For instance, we
associate the remarkable properties of supercooled liquids as they become
glassy
with the behavior of a uniformly frustrated system in the temperature range
$T\lesssim T_1$.  The structural correlation functions of certain high
temperature superconducting materials also exhibit behavior which is strikingly
similar to that of the various low temperature regimes of the present model.
It seems that avoided critical behavior is surprisingly robust, at least in $d
\le 3$.  It occurs naturally in the continuum version of the model, and
survives
lattice effects.  Preliminary results from the $1/n$ expansion\cite{zohar} show
that it persists at finite $n$. This is a new construct which
is likely to find applications in a variety of other arenas.

\subsection{Effects of Quantum Fluctations }

In the quantum model, there are two distinct types of possible ordered states:
charge-ordered states, in which $S_{\vec R}$ develops a non-zero
expectation value, and superconducting states, in which $P_{\vec R}$
develops a non-zero expectation value.  At zero temperature,
the ground state will be ordered if $s$
exceeds a parameter-dependent critical value $s_c$
while the ground state
is quantum disordered if $s < s_c$.
(When
factors of $\hbar$ are restored, $s^2\rightarrow s^2/\hbar$,
in Eq. (\ref{eq:quconstraint}) so $s^2$ is the natural
measure of the importance of
quantum fluctuations.)  The superconducting state can occur
only for $K$ greater than a critical value, $K_0$,
while the charge-ordered state
can occur only for $K < K_0$;
coexistence, {\it i.e.} supersolid order,
can occur only for the special case of $K=K_0$.
( For $Q \le 16$, $K_0=-2\sqrt{Q}$;  more generally, $K_0$ is that
value of $K$ for which the minimum over all $\vec k$ of ${\cal J}(\vec k)$
in Eq. (\ref{eq:calJtot}) is equal to $0$.)
As in the classical model, the transverse
quantum fluctuations in the continuum approximation (or for
$\lambda=0$) are sufficient to
destroy any possible charge order for small $Q$, even at zero temperature.
However, for non-zero $\lambda$, not only
is there a finite value of $s_c$ for $d > 1$, but it
is a continuous function of $Q$, even in the limit $Q \rightarrow 0$.  Indeed,
$s_c$ is a rather weak function of all parameters, and is
always greater than, but approximately equal to 1.  (The dependence of
$s_c$ on
$K$ and $Q$ for $W=1$
is shown in Fig. 4.)  Thus, at zero temperature and for $s > s_c$,
there is a superconducting
to charge-density wave transition
that occurs as a function of $K$.  Similarly, as a function
of decreasing $s$,
there is a superconducting to
quantum-disordered transition which occurs for $K>K_0$,
and a similar charge-density wave to quantum-disordered phase
transition
for $K<K_0$ .

At finite temperatures, the behavior of the system for
$K<K_0$ and $s > s_c$ is similar to that of the classical frustrated model
described above, while for $K>K_0$, the superconducting to normal
transition occurs in qualitatively the same way as in the ordinary
ferromagnetic spherical model.  However, so long as $|K-K_0|$ is small, both
the normal and superconducting states will exhibit substantial correlations
which resemble the nearby charge-density wave phase.
The same two crossover temperatures that appear in the correlation function
of the classical model calculations (Fig. 3), but in contrast to
that case, the spin
correlation length saturates at a long but finite value
for $T \le T_c$ where $T_c$ is the superconducting transition
temperature.

\subsection{Effects of Disorder }

For $Q>0$, disorder destroys the
possiblity of a charge-ordered state
in all dimensions $d\le 4$. For $K< K_0$,
the ground-state in $d\le 4$ has more or less
extended short-ranged
charge correlations, depending the strength of the disorder,
but no true long-range order.   (Indeed, the behavior of the disordered
system in dimension $d$ is qualitatively similar to that of the ordered
system in dimension $d-2$, or in other words the disorder
produces the standard ``dimensional reduction'' associated
with random field problems\cite{randomfield})
For $Q=0$, however, the fact that
the mean-squared random field, $f(\vec k)$, vanishes at small $k$ insures
that a finite transition temperature to a charge-ordered
(ferromagnetic) state survives up to a critical magnitude of the
disorder.   Thus, in all dimensions $d\le 4$,
the model has
an avoided critical point.  By contrast, weak disorder has relatively
little effect on the ground state and low-temperature properties of
the superconducting state unless $K-K_0$ is quite small.  For fixed
disorder, and $K>K_0$, the superconducting $T_c$ tends continuously
to zero as $K$ is decreased, and always vanishes before $K > K_0$.
(Recall that, in the absence of disorder,  the
superconducting state would have given way to a charge-ordered state, both
with finite transition temperatures,  at precisely $K=K_0$.)

In Fig. 5,
we show the phase diagram as a function of temperature and position along
the representative trajectory
through parameter space
indicated by the arrow in Fig. 4, both in the
presence and absence of disorder.

\subsection{Some Exact Ground-states for the Ising Model}

An amusing side benefit of the present analysis is that, for certain ranges of
parameters, the exact ground states of the classical spherical model
are also the exact ground states for all $n$ of the $O(n)$ model
with the same Hamiltonian, including the $n=1$ Ising case.
(These are
listed in Table I.)
The proof is simple:  For the spherical model, the ground states are
found by minimizing the Hamiltonian with respect to spin configurations,
subject
to one global constraint.  For the Ising model,
the same minimization problem must be solved, but now
subject to  $N-1$ additional constraints to insure that
each spin has length 1.  Thus, the ground-state energy of the Ising model must
always be greater than or equal to that of the spherical model.  If it so
happens that the ground state of the spherical model is an Ising state,
then this state must be a ground state of the corresponding Ising model.
To extend the proof to the $O(n)$ model, we simply consider $n$ copies
of the spherical model with one coupled constraint.

In this way, we can prove that there exists a {\it region} of
parameter space for which the corresponding frustrated $O(n)$
model has the spherical model ground state
(and is thus independent of $n$).  In three dimensions
these states are:  1) a six-fold
degenerate width 1 stripe
with $\vec q=(\pi,0,0)$,  2) a six-fold degenerate columnar state with
$\vec q=(\pi,\pi,0)$, and 3) a two-fold degenerate N\`eel state with
$\vec q = (\pi,\pi,\pi)$.  (The degeneracies apply to the Ising model;
for $n > 1$, the $O(n)$ models have an additional continuous degeneracy.)
In addition, it is possible to prove that there
exists at least a {\it surface} in parameter space on
which the ground-state is:
1) a 12-fold degenerate width 2 stripe with $\vec q=(\pi/2,0,0)$,
2) a 24-fold degenerate rectangular columnar state with $\vec q=(\pi,\pi/2,0)$,
3) a 12-fold degenerate rectangular N\`eel state with $\vec q=(\pi,\pi,\pi/2)$,
and 4) a 16-fold degenerate
width 2 N\`eel state with $\vec q =(\pi/2,\pi/2,\pi/2)$.
(Surely,
for the Ising model there is a commensurability
energy which
stabilizes these latter states in a finite region of parameter space.)
Similar results can straightforwardly
be obtained in other dimensions.

\section{Method of Solution}

\subsection{ The Classical Model}

Consider a model of the
form of Equation (\ref{eq:Hcl})
where for convenience, we add a counter-term
as in Eq. (\ref{eq:defK}) defined so
that the minimum value of ${\cal J}$ is zero.
The standard version \cite {BK} of the spherical
model dictates that we integrate over all configurations $(S_k)$ subject to
the constraint Eq. (\ref{eq:constraint}).  A  simpler solution, that was
introduced as early as 1952\cite {lw}, employs
the method of Lagrange multipliers:  The
original Hamiltonian is augmented by the term $\frac12\mu\sum_k|S_k|^2$ and the
integration takes place over {\it all } finite energy spin configurations.
The model is now unconstrained and quadratic, so all quantities can be computed
readily.  As long as $\mu > 0$, this can be done
without apologies and equation of constraint  becomes
an implicit equation for $\mu(T)$:
\be
\frac1T = \Phi_N(\mu) \equiv \frac1N\sum_{\vec k}\frac{1}{{\cal J}(\vec k) +
\mu}
\ee
or, in the thermodynamic limit,
\be
\frac1T = \Phi(\mu) = \int \frac{d^d k}{(2\pi)^d} \frac 1
{{\cal J}(\vec k) + \mu},
\label{eq:muofT}
\ee
and the integral is over the first Brillouin zone.
If this equation cannot be satisfied for any value of $\mu$, we are at or below
criticality.  Since $\Phi$ is a monotonically decreasing function of $\mu$,
$T_c$ is determined according to
\be
\frac1 {T_c} = \Phi(0^+).
\label{eq:Tcclassical}
\ee
For $T < T_c$, the Lagrange multiplier is set to zero.  The total population of
the finite
modes therefore is deficient, and the
remainder is identified as a condensate to be
distributed among the zero modes.

It is intuitively
clear that at least as far as the equilibrium properties of local
observables are concerned,
the above procedure is equivalent, in the thermodynamic limit, to
the original constrained model.  Such
results have been established with a
large degree of generality -- more
than sufficient to cover the cases of interest here.
See, for example, \cite{GP} (in particular Theorem 1) and references therein.

	Most of the claims
made in the earlier sections thus amount to explicit calculations or
elementary analysis of the
function ${\cal J}(\vec k)$.  In particular,
the internal energy per site $U$ as a function of temperature is simply
\be
U= \frac 1 2 \left[T - \mu(T) \right ]
\ee
and all other thermodynamic quantities can be determined by taking
appropriate partial derivatives of this expression.
For $T < T_c$, the magnitude, $m$, of the condensate is simply
\be
m^2= 1 - T \Phi(0) = [T_c-T]/T_c.
\ee
The spin--spin correlation functions
can be straightforwardly calculated according to
\be
\tilde G(\vec k, T)  \equiv
\langle |S_k|^2  \rangle = T/[{\cal J}(\vec k) + \mu(T)].
\label{eq:Gofk}
\ee
as is easily seen using the so called {\it method of generating
functions:} Add small terms to the
Hamiltonian (well localized in position or momentum
space) and
differentiate the free energy with respect to the appropriate coupling.
{}From Equation (\ref{eq:Gofk}),
the inverse of the correlation length may be extracted:
\be
\xi_a(T)^{-1} =\text{min}\left|\text{Im}\{q_a^0\}\right|
\label{eq:xi_a}
\ee
(in general, ``$\xi$'' may be anisotropic, hence it is labeled by a direction)
where $\vec q^{\thinspace0}$ is the solution of the implicit equation
\be
{{\cal J}}(\vec q^{\thinspace0}) + \mu(T) = 0,
\ee
and the minimization in Eq. (\ref{eq:xi_a}) is over the set of solutions
with all other components of $\vec q$ real.
(In general, ``$\xi$'' may be anisotropic, hence it is labeled by a direction.)

Crossover temperatures can, of course, never be determined from a
single sharp criterion.
However, near the avoided critical point, a
crossover occurs in a narrow range of temperatures, so it is both
useful and appropriate to define a crossover temperature explicitly.  The
crossover temperature,
$T_1$ (below which the frustration becomes ``important'') is therefore
defined as the solution of the implicit equation
\be
\mu(T_1) = 2\sqrt{Q}.
\ee
where the factor of 2 is chosen for aesthetic reasons.  Notice that for
an avoided critical point, $T_1 \to T_c(Q=0)$ as $Q\to 0$.  Along a
trajectory in the $Q - T$ plane which lies at fixed distance above $T_1(Q)$,
the correlation length approaches a finite limit as $Q\to 0$, while below
$T_1$, the correlation length diverges in this limit.
The lower crossover temperature $T_2$ (below which the lattice anisotropy
becomes ``important'')  will be defined -- with the same
degree of arbitrariness as $T_1$ -- as the solution of
\be
\mu(T_2)=\lambda Q.
\ee
For small $Q$ and $\lambda$ and $T_1 \gg T \gg T_2$,
the structure factor is dominated by a sharp ridge at $|\vec k|=q$,
while for $T_2 > T$, there are six sharp peaks at $|\vec k|=q$ and
$\vec k$ along a coordinate axis.  With our definition, at $T=T_2$, the
modulation in the magnitude of $S(\vec k)$ (as a function of angle)
with fixed $|\vec k|=q$, is comparable to the height of the ridge.

\subsection{The Quantum Model}

The quantum model is solved in much the same way as the classical model,
through the introduction of a chemical potential to enforce the constraint,
which reduces the problem to a set of decoupled harmonic oscillators.
Thus, for all temperatures
above $T_c$, $\mu$ is implicitly determined from the relation
\be
s^2=\Phi_{qu}(\mu,T)
\label{eq:muquofT}
\ee
where
\be
\Phi_{qu}(\mu,T)=\frac 1 2 \int \frac{d^dk} {(2\pi)^d} \left\{\left[
\sqrt{W\Delta(\vec k)+\mu\over {\cal J}(\vec k) + \mu} +
\sqrt{ {\cal J}(\vec k) + \mu \over W\Delta(\vec k)+\mu }\right]\left[
2n(\omega_{\vec k}/T) + 1\right]\right\},
\ee
\be
\omega_{\vec k} = \sqrt{\left[ W\Delta(\vec k)+\mu\right]
\left[{\cal J}(\vec k) + \mu \right]},
\ee
and $n(x)=\left[e^x-1\right]^{-1}$ is the Bose occupation factor.
(The behavior of the dispersion relation as $k\to 0$ is somewhat
peculiar, but this has no effect on the present results.\cite{plasma})
Again,
$\Phi$ is a monotonically decreasing function of $\mu$ so, at $T=0$,
the critical value of $s$
is determined from the equation
\be
s_c^2=\Phi(0,0)
\ee
and, when $s>s_c$, the critical temperature $T_c$ is determined
according to
\be
s^2=\Phi(0,T_c).
\label{eq:Tcqu}
\ee
For $s > s_c$ and $T < T_c$,
the magnitude of the condensate, $m$, is determined according
to
\be
m^2= s^2 - \Phi(0,T),
\ee
where, as discussed above, $m$ is the magnitude of the superconducting
condensate for $K > K_0$ and the amplitude of the charge-density wave
for $K < K_0$.
The dynamic spin correlation function (since in quantum mechanics,
dynamics and thermodynamics are intimately related, the Fourier transform
of the two-time spin correlation function is the fundamental
quantity) is easily seen to be
\be
\tilde G(\vec k, \omega) =\frac 1 2
\sqrt{W\Delta(\vec k)+\mu\over {\cal J}(\vec k) + \mu}
\left\{
\left[n(\omega_{\vec k}/T) + 1\right]\delta(\omega-\omega_{\vec k})
+ n(\omega_{\vec k}/T)\delta(\omega+\omega_{\vec k})\right\},
\label{eq:Gqu}
\ee
while the static structure factor,
\be
\tilde G(\vec k)=\int \frac {d\omega} {2\pi}
\tilde G(\vec k, \omega) =\frac 1 2
\sqrt{W\Delta(\vec k)+\mu\over {\cal J}(\vec k) + \mu}
\left[2n(\omega_{\vec k}/T) + 1\right].
\ee
The inverse charge-ordering correlation length
(which, again, will generally be
anisotropic)
is
\be
1/\xi_a^{ch}(T) =\text{min}\left|{\rm Im}\{q_a^{ch}\}\right|
\ee
where $\vec q^{ch}$ is the solution of the implicit equation
\be
{\cal J}(\vec q^{ch}) + \mu(T) = 0,
\ee
and the minimization is again performed with all other components
of $\vec k$ real.  A similar expression for the
superconducting correlation function is easily obtained by
inverting the term in the square root in Eq. (\ref{eq:Gqu})
and a superconducting correlation length obtained from the
solution of the implicit equation
\be
W\Delta(\vec q^{sc}) + \mu(T) = 0.
\ee
Finally, the internal energy per site
$U$ as a function of temperature is simply
\be
U= U_o(T) - \frac {s^2} 2 \mu(T)
\ee
where
\be
U_o(T)= \frac 1 2 \int {d^dk\over (2\pi)^d} \omega_{\vec k}
\left[2n(\omega_{\vec k}/T) + 1\right]
\ee
is the internal energy of independent harmonic oscillators,
and all other thermodynamic quantities can be determined by taking
appropriate partial derivatives of this expression.

\subsection{The Model with Disorder}

Because of the harmonic nature of the model, the effect of an arbitrary
configuration of random fields $\{h_j\}$ can be formally accounted for
by shifting all spins according to
\be
\tilde S_{\vec k} \rightarrow \tilde S_{\vec k} + \tilde h_{\vec k}/
\left[ {\cal J}(\vec k) + \mu\right]
\ee
for either the classical or the quantum model.  The result is an
additive term to the internal energy,
\be
U \rightarrow U - \frac 1 2 \int {d^dk\over (2\pi)^d}
{\left|\tilde h_{\vec k}\right|^2\over
\left[ {\cal J}(\vec k) + \mu\right]},
\ee
and a shift in the implicit equation for $\mu$ brought about by the
substitution
\be
\left<\left|\tilde S_{\vec k}\right|^2\right > \rightarrow
\left<\left|\tilde S_{\vec k}\right|^2\right > +
{\left<\left|\tilde h_{\vec k}\right|^2\right >\over
\left[ {\cal J}(\vec k) + \mu \right]^2}.
\ee
Upon configuration averaging, this relation implies
\be
\Phi_{dis}(\mu,T)=\Phi_{qu}(\mu,T) + \int {d^dk\over (2\pi)^d}
{f(\vec k)\over
\left[ {\cal J}(\vec k) + \mu\right]^2},
\label{eq:Fdis}
\ee
(and the obvious corresponding equation for the classical model).
The spin correlation function in the presence of disorder is altered both
by the implicit change produced by the altered temperature dependence of
$\mu$, and by the addion of a zero fequency additive contribution
\be
\tilde G_{dis}(\vec k, \omega)= \tilde G(\vec k,\omega) +
{\delta(\omega)f(\vec k)\over
\left[ {\cal J}(\vec k) + \mu\right]^2}.
\ee
It is clear from Eq. (\ref{eq:Fdis}) that for $K-K_0$
negative $\Phi_{dis}$ diverges as
$\mu \rightarrow -K_0$ for all dimensions $d\le 4$, so that no
charge ordered state is possible.  Moreover, if $K-K_0$ is positive but small,
$\Phi_{dis}(0,T)$ will have a large, additive contribution from the
disorder term, from which it follows that the superconducting $T_c$ must
always vanish as $K\to K_0+\Delta K$  where $\Delta K$ is a positive,
increasing function of the strength of the disorder.

\section{ Mathematical Dump}

We now provide the calculational details
which underly the various claims made in the previous sections.
Our starting point is an elementary calculation
concerning the stationary points of ${\cal J}(\vec k)$:

\noindent{\bf Proposition IV.1.}
{\it Let
\be
{\cal J}_Q(\vec k) = \sum_aV(k_a) + \frac12\lambda \sum_{a\neq b} V(k_a)V(k_b)
+ \frac
Q{\sum_aV(k_a)} + K
\label{eq:JQ}
\ee
with $V(k_a) = 2(1-\cos k_a)$ and $\lambda \neq 0$.
Let $\vec p$ be a stationary point of ${\cal J}(\vec k)$ in the
first Brillouin zone, $-\pi < k_a \le \pi$.
Then any component of $\vec p$ may be $0$ or $\pi$
but the components that are not $0$ or $\pi$ are equal in magnitude
to each other.
In particular, if
$\vec p$ is a stationary
point then $|p_a| = 0$, $\pi$, or $q_{\ell, s}$ with $q_{\ell, s}$
satisfying
\be
1 + \lambda \left[4\ell + (s-1) V(q_{\ell, s})\right]
 = \frac{Q}{(4\ell + sV(q_{\ell, s}))^2}
\label{eq:f}
\ee
where $\ell$ is the number of
components equal to $\pi$ and $s$ is the number of the remaining
components that are non--zero.  This theorem applies in arbitary
dimension, $d\ge 1$.}

\noindent{\it Proof.}
Upon differentiating ${\cal J}(\vec k)$ we find
\be
[1 + \lambda\sum_{b\neq a} V(k_b) - \frac{Q}{[\sum_b V(k_b)]^2}]\sin k_a = 0.
\label{eq:stationary}
\ee
This is obviously satisfied if $k_a$ = $0$ or $\pi$.  Now suppose that
$\vec k$ satisfies Eq. (\ref{eq:stationary})
with two or
more components, $k_b$ and $k_c$ not $0$ or $\pi$.  Then, subtracting, we
get
\be
\lambda V(k_b) = \lambda V(k_c)
\ee
which implies $|k_b| = |k_c|$.  If there are
$\ell$ directions where $|k_a| = \pi$ and $s$ directions where $|k_a|$ is not
$0$ or $\pi$
then the magnitude of these remaining components satisfies the stated equation.
$\Box$

	What follows is the starting point for both the analysis of the low
temperature behavior in
the system as $Q\to 0$ and for the analysis of the ground state space.  For the
most part,
the indices
on the $q$ defined above will be understood from context and omitted.

\noindent{\bf Proposition IV.2}
{\it Let ${\cal J}(\vec k)$ be as described in Proposition IV.1.  Then for
$\lambda > 0$, and $Q
\leq 16$, the minimizing wave vector has a single non--zero component of
magnitude $q$
satisfying}
\be
V(q) = \sqrt Q.
\ee

\noindent
{\it Proof.}  If $\lambda$ = 0, the function has an absolute minimum of $2\sqrt
Q$ which is
achieved if $\sum_a V(k_a) = \sqrt Q$.  Provided that $\sqrt Q \leq 4$, this
value can be
obtained even if $\lambda > 0$ by a vector with only a single component that
has the above
stated magnitude.
$\Box$

We are now ready for our principal result for this section:

\noindent{\bf Theorem IV.3.}
{\it In three dimensions,
for the model described in Proposition IV.1, there is an
avoided critical point at $Q=0$.  In particular, for fixed $\lambda > 0$,
let $T_c(0)$
$\ (= T_c(0;\lambda))$
be the critical temperature for the $(Q = 0)$, ferromagnetic
version of the model:
\be
\frac 1{T_c(0)} = \frac{1}{(2\pi)^3}\int_{|k_a| < \pi}
\frac{d^3k}{\sum_aV(k_a) + \frac12\lambda\sum_{a\neq b}V(k_a)V(k_b)}
\ee
and $T_c(Q)$ given as in Eqs. (\ref{eq:muofT}) and
(\ref{eq:Tcclassical}) with
${\cal J}_Q(\vec k)$ given in Eq. (\ref{eq:JQ}) with $K=K_0=-2\sqrt{Q}$.
Then
\be
\frac 1{T_c(0)} < \lim_{Q\to0} \frac 1{T_c(Q)} = \frac 1{T_c(0)}+\frac
{B} {\sqrt\lambda },
\label{eq:TcQ}
\ee
Where $B = \frac{1}{(16\pi^2)}\int
d\Omega[\sin^4\theta
\cos^2\phi\sin^2\phi + \cos^2\theta\sin^2\theta]^{-\frac 12}
\approx 0.14245$}

\noindent
{\it Proof.}
For $Q \ll 1$, let
$q \simeq Q^\frac14$ denote the solution of
\be
V(q) = \sqrt Q
\ee
and let $\Delta$ be a small number independent of $Q$, the precise
specifications of which will be detailed later.  (In most of what is to follow,
the
distinction between $q$ and $Q^\frac14$ is practically irrelevant -- $q$ may
simply be
regarded as convenient notation for
$Q^\frac14$.)  It is clear that for the large-$k$ portion of the
integral,
$|\vec k| >
\Delta$, as $Q\to0$, nothing particularly interesting happens to the integrand
$[{\cal J}^{-1}_Q(\vec k)]$. Hence
\be
\lim_{Q\to 0}\int_{|\vec k| > \Delta} \frac{d^3k}{{\cal J}_Q(\vec k)} =
\int_{|\vec k| > \Delta} \frac{d^3k}{{\cal J}_{Q=0}(\vec k)}.
\label{eq:largek}
\ee
Notice that,
for small $\Delta$, the right hand side is only just shy of $1/T_c(0)$.

Now, let $0< t < 1$ and define the two constants $A_{\pm}=1\pm q^t$,
which have the property that $A_{\pm} \to 1$ as $Q \to 0$.
In terms of these, we
break the remaining region of integration in three:  1)  A
small $k$ region, ${\cal R}_1$
with $0 < |k| \le q A_-$;  2)  A critical region, ${\cal R}_2$ with
$q A_- < |k| \le  qA_+$;  3)  An intermediate $k$ region
${\cal R}_3$ with
$q A_+ < |k| \le \Delta$.  We shall show that, as $Q\to 0$,
the contribution from region
1 vanishes, the contibution from region 3 plus the contribution
from large $k$ in Eq. (\ref{eq:largek}) converge to $1/T_c(0)$,
and the contribution from the critical region approaches $
B/ {\sqrt\lambda }$.

\noindent{$\bullet$}
In ${\cal R}_1$,
as an upper bound, we will neglect the $\lambda$--perturbation in the
denominator.  Noting that in the specified region,
$\sum_a V(k_a) + \frac {Q}{\sum_a V(k_a)}$ is increasing and that, in general,
$V(k_a) \leq k_a^2$, we have
\be
\int_{{\cal R}_1}\frac{d^3k}{{\cal J}_Q(\vec k)}\leq
\int_{{\cal R}_1}\frac{d^3k}{k^2 + Q/k^2 - 2\sqrt Q}\leq
4\pi q^4\int_{|k| < qA_-}\frac{dk}{(k - Q^{\frac 14})^2}.
\ee
The last term is bounded by constants times $q^{(3 - t)}$ as $Q\to 0$
and thus the limiting contribution from this region may be neglected.

\noindent{$\bullet$}
Next, consider the intermediate region ${\cal R}_3$.
Since ${\cal J}_Q = {\cal J}_0 + Q/\sum_aV(k_a) - 2 \sqrt Q$ it follows that
\be
\int_{{{\cal R}}_3} \frac{d^3k}{{\cal J}_Q(\vec k)} =
\int_{{{\cal R}}_3} \frac{d^3k}{{\cal J}_0(\vec k)} +
\int_{{{\cal R}}_3} \frac{[2 \sqrt Q - Q/\sum_aV(k_a) ]d^3k}{{\cal J}_0(\vec k)
{\cal J}_Q(\vec k)}.
\label{eq:intermedk}
\ee
Now the first term on the right hand side of Eq. (\ref{eq:intermedk})
may be added directly to the
right hand side of Eq. (\ref{eq:largek})
to obtain, in the small $Q$ limit, exactly $1/T_c(Q=0)$.  Let
us show that the second term vanishes as $Q \to 0$.
In this region, the numerator is positive so we may discard the term involving
$\sum_a V(k_a)$.
The denominator is made smaller if we set $\lambda = 0$ so let us do that as
well.  Let $D$ denote any positive constant for which $V(k_a) \geq Dk_a^2$
holds
for all $k_a$ with $|k_a| \leq \pi$.  Putting these
together, and canceling powers of $k$ the
task is to show that
\be
\sqrt
Q\int_{{{\cal R}}_3}\frac{d^3k}
{\sum_aV(k_a) - 2\sqrt Q + \frac{Q}{\sum_aV(k_a)}}
\ee
vanishes in the $Q\to 0$ limit.
For the term $\frac{Q}{\sum_aV(k_a)}$ in the denominator,
we may use the bound $V(k_a) \leq
k_a^2$ and,
further, for the other appearance of ${\sum_aV(k_a)}$, we may replace this sum
with $k^2 - Ek^4$
where $E$ is some constant independent of $Q$ (or $\Delta$).  With these
estimates in tow,
the problem is one dimensional and, were it not for the quartic term,
would be entirely trivial.  In any case, we are now reduced to showing that
\be
\int_{{{\cal R}}_3}\frac{dk \sqrt Q k^2}{(k^2 - \sqrt Q)^2 - Ek^6}
\ee
vanishes as $Q\to 0$

Let us now
assert that $\Delta$ has been chosen small enough so that (for all $q$
sufficiently small) throughout the range $qA_+  \leq k \leq \Delta$,
\be
Ek^6 < \frac12 (k^2 - \sqrt Q)^2.
\ee
Indeed, provided that $q$ is sufficiently small,
the inequality clearly holds at the lower
limit and,
upon comparison of derivatives, the desired inequality holds throughout the
entire range provided
that $E\Delta$ is somewhat less than one.  What is left after these
estimates can, essentially, be done by hand:
\be
\int_{{{\cal R}}_3}\frac{\sqrt Q k^2dk}{(k^2 - \sqrt Q)}
\leq \int_{{{\cal R}}_3}\frac{\sqrt Qdk}{(k - Q^{\frac14})^2}
\leq\frac{\sqrt Q}{q -Q^{\frac14} + q^{(1 + t)}}
\approx q^{(1 - t)}
\ee
which indeed vanishes as $Q\to 0$.

\noindent{$\bullet$}
Thus we are left with the critical region ${\cal R}_2$.  In this region,
the deviation of
$k$ from $q$ is so slight that we are essentially in the
position where we can ``expand
and neglect''.
In particular, upper and lower bounds may be derived, in this region,
by setting various items
to their maximum or minimum value.
Since the procedure is
similar on
both sides, we will be content with an explicit derivation of a lower bound
on the remaining
integral that agrees with the stated formula.  As will become apparent,
an upper bound can be derived in the same fashion.

In what follows, all
constants
$C_n$ will be functions of $Q$ with the property that $C_n\to 1$ as $Q\to 0$.
It is slightly easier to work with the quantities $V(k_a)$ instead of the $k_a$
hence we define the variables
\be
v_a = [V(k_a)]^{\frac12}
\ee
and note that $dv_a = |\frac{\sin k_a}{2\sin(\frac 12 k_a)}|\geq C_1dv_a$.
It is further clear that in the region $|k-q| < q^{1 + t}$, we may write
$d^3k \geq C_2q^2dvd\Omega$.  Next, it is noted that the image of the region
$|k-q| < q^{1 + t}$, in $v$--space, also contains a spherical shell of size
$C_3q^{(1 + t)}$.  We will
confine our attentions to this smaller region.
Thus,
\be
\int_{{\cal R}_2} \frac {d^3k} {{\cal J}(\vec k)} \geq C_2 q^2\int_{|u| <
C_3q^{1+t} } \frac {du d\Omega} {4C_4u^2 + \lambda C_5 q^4 R(\Omega)},
\ee
where $u=v-q$,
\be
R(\Omega) = \lim_{q\to 0}\frac 1{q^4}\frac12\sum_{a\neq b}v_a^2v_b^2
\equiv [\sin^2\theta\cos^2\theta + \sin^4\theta
\cos^2\phi\sin^2\phi].
\ee
For $R(\Omega)\ne 0$, the $u$ integral is easily preformed to yield
\be
\int_{{\cal R}_2} \frac {d^3k} {{\cal J}(\vec k)} \geq
\frac {C_5} {\sqrt{\lambda}} \int
\frac {d\Omega \alpha_o} {\sqrt{ R(\Omega)} },
\ee
where
\be
\alpha_o=\tan^{-1}\left[2C_3q^{t-1}/\sqrt{\lambda R(\Omega)} \right]\to \pi/2.
\ee
Hence, it is seen that
\be
\lim_{Q\to 0}
\int_{{\cal R}_2} \frac {d^3k} {{\cal J}(\vec k)}
\geq
\frac{\pi}{2\sqrt{\lambda}}\int\frac{d\Omega}{\sqrt{R(\Omega)}}
\ee
as claimed.

An upper
bound follows almost the identical derivation with a renaming of the constants
$C_n$.
$\Box$

\noindent{{\bf Corollary.}}
{\em  In dimension $d>3$, there is a ``nearly avoided critical
point'' for the stated model in the sense that for
fixed $\lambda > 0$, as $Q\to 0$,}
\be
\frac{1}{T_c(Q)} - \frac{1}{T_c(0)}
\sim Q^{\frac{d-3}{4}}\lambda^{-\frac 12}B_d
\ee
{\it where}
$B_d = (2\pi)^{-d}\int_\Omega d\Omega [R_d(\Omega)]^{-\frac 12}$
{\it with} $R_d(\Omega) = \frac 12 \sum_{a\neq b}x_ax_b\mid_{x^2 = 1}$.

\noindent {\it Proof.}
If we use the same division of regions and follow step for step
the analysis of
Theorem IV.3, we see that all the estimated quantities that go to zero
are now multiplied by an extra factor of $k^{(d-3)}$ (from the volume element).
This
always amounts to an extra factor of $q^{(d-3)}$  Similarly, the term that was
of
primary interest works in a similar way but is multiplied by  $q^{(d-3)}$.
Everything else
that was finite produces
$\frac{1}{T_c(0)}$ mutatis--mutandis.  In particular, we get
\be
\lim_{Q\to 0}Q^{-\frac {d-3}{4}}[\frac{1}{T_c(Q)} - \frac{1}{T_c(0)}]
= \lambda^{-\frac 12}B_d
\ee
$\Box$

Our discussion of the ground state space now picks up where Proposition IV.2
left off.  Right now, for $\lambda > 0$ and $Q < 16$ the situation
is well under control and these are the sorts of results that we seek
throughout the phase plane.  For the sake of brevity, we will focus our
attention on the cases of principal interest, namely $d = 3$ and $\lambda >
0$.

Proposition IV.1 tells us that in any region, there are only a finite number
of possibilities to consider; however, for $d = 3$, this turns out to be eight
additional distinct modes (other than $(q,0,0)$) and five new regions.
Notwithstanding, we will attempt to be as brief as possible and still lay claim
to a rigorous proof; this is most efficiently carried out by writing out five
separate sub--propositions.  A complete list of the
competing modes as well as the regions of interest can be found in Fig. 2.
In the up and coming, the various state (or modes) will not be
distinguished from their reflection or coordinate axis exchange equivalents.
Thus, e.g. Proposition IV.4.i below really pertains to six ground (equivalent)
states.

\noindent {\bf Proposition IV.4.i.\ }
{\it In the region $\lambda > 0$, $Q > 16$ and $Q < 16(1 + 4\lambda)$, the
ground
state is of the form $(\pi,0,0)$}.

\noindent {\it Proof.\ }  Since $Q > 16$, the
$(q,0,0)$ states need not be considered because there is no solution to
Eq. (\ref{eq:f})
for $q$. Similarly, for the mode $(\pi,q,0)$, the defining equation
reads $(1 + 4V_q)^2 = \frac{Q}{1 + 4\lambda}$ (where here, and in what is to
follow, we use the notation $V_q \equiv V(q)$).  Hence this mode is forbidden
if $Q < 16(1 + 4\lambda)$.  In the same region, for the same reason,
the mode $(\pi,q,q)$ is disallowed and similarly, the mode
$(\pi,\pi,q)$ is forbidden for
$Q < 64(1 + 8\lambda)$.  Now the ``energy'' for the current state,
${\cal J}(\pi,0,0)$, is simply
$4 + Q/4$.  In the stated region, this is a whole lot less than $8 + Q/8 +
16\lambda$ and
$12 + Q/12 + 48\lambda$ which eliminates $(\pi,\pi,0)$ and $(\pi,\pi,\pi)$ from
consideration.  This leaves as contenders only the modes $(q,q,q)$ and
$(q,q,0)$.  The energy for $(q,q,0)$ is given by
\be
{\cal J}_(q,q,0)
\equiv {\cal E}_{q,q,0} = 2V_q + \frac{Q}{2V_q} + \lambda V_q^2.
\ee
Now in general, the energy of any state increases with
$\lambda$ however ${\cal E}_{\pi,0,0}$ is independent of $\lambda$.  It is
therefore sufficient to establish ${\cal E}_{q,q,0} > {\cal E}_{\pi,0,0}$ at
the
lower--right boundary of the region under consideration.  To this end, we write
\be
  {\cal E}_{q,q,0}
= 2V_q + \frac{Q}{2V_q} + 4\lambda[2V_q] - 16\lambda + \lambda(4-V_q)^2.
\label{eq:qq0}
\ee
Neglecting the quadratic term and minimizing at $Q = 16(1 + 4\lambda)$, this is
maximized when $2V_q = 4$ and weighs in at exactly $4 + Q/4$.  Using the fact
that we are
interested in $Q$'s that are {\it strictly} less than
$16(1 + 4\lambda)$ and, the fact that ${\cal E}_{q,q,0}$ is a strictly
increasing function of
$\lambda$ we find ${\cal E}_{q,q,0} > {\cal E}_{\pi,0,0}$.  The mode $(q,q,q)$
has
energy $3V_q + Q/3V_q + 3\lambda V_q^2$ which (when regarded as a function of
$3V_q$)
is manifestly larger than ${\cal E}_{q,q,0}$.

\noindent {\bf Proposition IV.4.ii.\ }
{\it In the region $\lambda > 0$,  $16(1 + 4\lambda) < Q < 64(1 + 4\lambda)$,
the
ground state is of the form $(\pi,q,0)$.}

\noindent {\it Proof.\ }
Notice that this is exactly the region where Eq. (\ref{eq:f}) has a
solution for the mode $(\pi,q,0)$.  The mode $(\pi,\pi,q)$ has no solution in
this region (and nor does $(q,0,0)$).  We write
\be
 {\cal E}_{\pi,q,0} = \min_\omega[\omega +
\frac Q\omega + 4\lambda\omega - 16\lambda]
\label{eq:piq0}
\ee
(with $\omega \equiv 4+V_q$ ; note that the minimizing $\omega$ equals
$\sqrt{Q/(1 + 4\lambda)})$.
On comparison to ${\cal E}_{q,q,0}$ as expressed in
Eq.(\ref{eq:qq0}),
it is clear that ${\cal E}_{\pi,q,0}$ is lower.  This also eliminates
$(q,q,q)$ on the basis of the final argument in Proposition III.4.i.
Next we have ${\cal E}_{\pi,\pi,0} = 8 + Q/8 + 16\lambda$ and plugging
$\omega = 8$ into the expression on the right-hand side of Eq.
(\ref{eq:piq0}),
this is
exactly what we get.  The minimizing $\omega$ will do better.

Thus we may turn our attention to the mode at $(\pi,q,q)$ -- the ones at
$(\pi,0,0)$ and $(\pi,\pi,\pi)$ then follow immediately.  The energy,
${\cal E}_{\pi,q,q}$ admits the expression
\be
{\cal E}_{\pi,q,q} = 4 + 2V_q + \frac{Q}{4 + 2V_q} +
4\lambda(4 + 2V_q) - 16\lambda + 8\lambda V_q^2.
\ee
Regarding this as a function of $4 + 2V_q$, and comparing to Eq.
(\ref{eq:piq0}),
this
energy is clearly larger than ${\cal E}_{\pi,q,0}$ whenever we are in a region
where a minimizing $\omega$ for Eq. (\ref{eq:piq0}) exists.

\noindent {\bf Proposition IV.4.iii.\ }
{\it In the region $\lambda > 0$,  $(1 + 4\lambda) < Q/64 < 64(1 + 8\lambda)$,
the
ground state is of the form $(\pi,\pi,0)$}.

\noindent{\it Proof.\ }  The modes $(q,0,0)$ and $(\pi,q,0)$ are eliminated
from
consideration.  Similarly, $V_q$ for $(q,q,0)$ must satisfy
\be
4V_q^2 = \frac Q{1 + \lambda V_q}
\ee
but
\be
\frac Q{1 + \lambda V_q} \geq \frac Q{1 + 4\lambda} > 64
\ee
so for $(q,q,0)$, $V_q$ cannot get big enough.

	For the mode at hand, the energy ${\cal E}_{\pi,\pi,0}$ is given by the simple
formula
\be
{\cal E}_{\pi,\pi,0} = 8 + Q/8 + 16\lambda.
\ee
Subtracting this from various expressions for the energy of various other
modes,
we see
${\cal E}_{\pi,0,0} - {\cal E}_{\pi,\pi,0}\propto Q - 16(1 + 4\lambda) > 0$,
${\cal E}_{\pi,\pi,\pi} - {\cal E}_{\pi,\pi,0} = 4(1 + 8\lambda) - Q/24 > 0$
and similarly,
${\cal E}_{\pi,\pi,q} - {\cal E}_{\pi,\pi,0} =
V_q(1 + 8\lambda -Q/(8)(8+V_q)) > 0$.
Notwithstanding the losing status of the state $(\pi,\pi,q)$, for the benefit
of
the final two states on the list, let us write its energy:
\be
{\cal E}_{\pi,\pi,q} = 8 + V_q + \frac{Q}{8 + V_q} + \lambda[16+8V_q]
= \eta + \frac Q \eta +8\lambda\eta -48\lambda
\label{eq:pipiq}
\ee
where $\eta \equiv 8 + V_q$.
Similarly, ${\cal E}_{q,q,q}$ may be written
\be
{\cal E}_{q,q,q} = \sigma + \frac Q\sigma + 8\lambda\sigma -48\lambda
+ \lambda[\frac13\sigma^2 + 48 -8\sigma]
\label{eq:qqq}
\ee
with $\sigma = 3V_q$.  (Of course the $q$'s referred to in Eqs.
(\ref{eq:pipiq}) and (\ref{eq:qqq})
pertain to different solutions of Eq. (\ref{eq:f})
and are not to be identified with one another.)
It is easily checked that $\frac13\sigma^2 + 48 -8\sigma > 0$ for $\sigma < 12$
and
hence it is seen that ${\cal E}_{q,q,q} > {\cal E}_{\pi,\pi,q}$.

Similarly, we write
\be
{\cal E}_{\pi,q,q} = 4 + 2V_q + \frac{Q}{4 + 2V_q} + \lambda(8V_q + V_q^2) =
\zeta + \frac{Q}{\zeta} + 8\lambda\zeta -48\lambda +
\lambda[\frac14 \zeta^2 - 6\zeta + 36]
\ee
and again the term in square brackets is positive if $\zeta < 12$.

\noindent {\bf Proposition IV.4.iv.\ }
In the region $\lambda > 0$ and $64 < Q/(1 + 8\lambda) < 144$, the
ground state is of the form $(\pi,\pi,q)$.

\noindent {\it Proof.\ }
The modes $(q,0,0)$, $(\pi,q,0)$ and $(q,q,0)$ are already out of the loop.  In
Proposition
IV.4.iii, we have (just) shown ${\cal E}_{q,q,q} > {\cal E}_{\pi,\pi,q}$ and
${\cal E}_{\pi,q,q} > {\cal E}_{\pi,\pi,q}$ so these are out.  Now at the
minimum,
$V_q$ satisfies $(8 + V_q)^2 = Q/(1 + 8\lambda)$.  In the specified region,
this is not solved
by $V_q = 0$ or $V_q = \pi$ so evidently ${\cal E}_{\pi,\pi,\pi} > {\cal
E}_{\pi,\pi,q}$ and
${\cal E}_{\pi,\pi,0} > {\cal E}_{\pi,\pi,q}$.  Finally, we dispense with
$(\pi,0,0)$ by noting
that in this region, ${\cal E}_{\pi,\pi,0} < {\cal E}_{\pi,0,0}$.

\noindent{\bf Proposition IV.4.v.\ }
{\it In the region $\lambda > 0$, $Q > 144(1 + 8\lambda)$ the
ground state is of the form $(\pi,\pi,\pi)$.}

\noindent {\it Proof.\ }
Examining Eq. (\ref{eq:f})
for $V_{q_1}$ in the state $(q_1,q_2,q_3)$, we have
\be
1 + \lambda(V_{q_2} + V_{q_3}) = \frac Q{(V_{q_1} + V_{q_2} + V_{q_3})^2}.
\ee
The right hand side is larger than $Q/144$ and the left hand side is smaller
than $(1 + 8\lambda)$.  Thus there are no solutions with any component not
equal to $0$ or
$\pi$.  It follows from earlier results (or it can be easily checked) that in
this region,
${\cal E}_{\pi,\pi,\pi}< {\cal E}_{\pi,\pi,0} < {\cal E}_{\pi,0,0}$.

$\Box$
\noindent{\bf Corollary.\ }  {\it Along the line $\frac {Q}{1+4\lambda} = 36$,
the ground state
is $(\pi,\frac{\pi}{2},0)$ and along the line
$\frac {Q}{1+8\lambda} = 100$, the ground state
is $(\pi,\pi,\frac{\pi}{2})$}

\noindent {\it Proof.\ }  This follows from setting $V_q = 2$ in the
appropriate
region and solving Eq. (\ref{eq:f}) for $Q(\lambda)$

$\Box$

\noindent {\it Remark}  The case $\lambda < 0$
is far easier to analyze.  Indeed, writing
${\cal J}(\vec k) = \Delta(\vec k) + \frac 12 |\lambda|[\Delta(\vec k)]
+ Q/\Delta(\vec k) - \frac 12|\lambda|\sum_aV(k_a)^2$, let $\vec k$ denote any
wave vector, and let $\tilde q$ satisfy $dV(\tilde q) = \Delta(\vec k)$.  It is
clear that ${\cal J}(\tilde q, \dots, \tilde q) < {\cal J}(\vec k)$ unless
$\vec k = ((\tilde q, \dots, \tilde q))$.  Indeed, this amounts to showing
that
$\sum_aV^2(k_a) \geq \frac 1d [\Delta(\vec k)]^2$ and the latter is just (the
discrete form of) H\"older's inequality which holds as an equality if and only
i
f
the $V(k_a)$ are independent of $a$.  Evidently the minimizer is ``diagonal''
fr
om which
it is easy to see that (in three dimensions) the N\'eel ground state dominates
in the region $Q \leq 144 - 1152|\lambda|$ for $\lambda < 0$.

\section{acknowlegements}
Much of this work was inspired and guided by the keen insights of D.~Kivelson.
We are grateful to J.~Rudnick and J.~P.~Sethna for useful discussions.
SK wuold like to aknowledge the hospitality of the Institute for Theoretical
Physics in Santa Barbara where much of this work was carried out, and LC
and SK express extreme appreciation of City Bean Coffee, where much
of this project was conceived.  This work was supported in part by the
National Science Foundation grant number DMR93-12606 (SK and ZN)
and DMS-93-02023 (LC) at UCLA and PHY94-07194 (SK) at UCSB.  Work at
Brookhaven (VE) was supported by the Division of Materials Science,
U. S. Department of Energy under contract No. DE-AC02-76CH00016.
\newpage

\vspace*{2cm}
\begin{center}
\begin{tabular}{|c|c|c|c|c|}
\hline
{$\lambda > 0$} & {$0 < Q < 16$} & {$\vec q = < q,0,0 >$} &
{$V(q)=\sqrt{Q}$} & \\
\hline
{$ \lambda > 0$} & {$16 \le Q \le 16(1+4\lambda$} &
{$\vec q = < \pi,0,0 >$} & { -- } &\\
\hline
{ $\lambda > 0$} & {$16(1+4\lambda) < Q < 64(1+4\lambda)$}
& {$\vec q = < \pi,q,0 >$} & {$V(q) = \sqrt{Q/(1+4\lambda)}-4$} & \\
\hline
{ $\lambda > 0$} & {$64(1+4\lambda) \le Q \le 64(1+8\lambda)$}
& {$\vec q = < \pi,\pi,0 >$} & { --  } & \\
\hline
{$\lambda > 0$} & {$64(1+8\lambda) < Q < 144(1+8\lambda)$} &
{$\vec q = < \pi,\pi,q >$} & {$V(q)=\sqrt{Q/(1+8\lambda)}-8$} & \\
\hline
{$ \lambda > 0$} & {$144(1+8\lambda) \le Q$} & {$\vec q = < \pi,\pi,\pi >$}
& { -- } & \\
\hline
{ $-1/8 < \lambda < 0$} & {$0 < Q < 144-1152\left|\lambda\right|$}
 & {$\vec q = < q,q,q >$} &
{complicated } & \\
\hline
{$ \lambda < 0$} & {$144-1152\left|\lambda\right|
\le Q $} & {$\vec q = < \pi,\pi,\pi >$} &
{ -- } & \\
\hline
\hline
\end{tabular}
\end{center}
Table 1: Values of the three dimensional wave vector,
$\vec q$, that minimize ${\cal J}(\vec k)$
in different ranges of $Q$.  This determines the locations of
the peaks in the structure factor in the disordered phase and
the location of the Bragg peaks in the charge-ordered phase.
The value of $q$ labeled ''complicated'' is given by
the solution of $V(q)=v$ where $v$ is the solution of the
cubic equation $9v^2=Q+18|\lambda|v^3$.
\vspace*{2cm}
\vspace*{1cm}
\section{Figures}

1) Phase diagram in  the $T$ - $Q$ plane for the classical model
in $d=3$.
\bigskip

2) Pictures of the various ordered phases of the Ising version of the
model for positive $\lambda$.
\bigskip

3) $\tilde G(\vec k)$ for Q=1/2 and $\lambda =1/4$
\bigskip

4) Contour plot of $s_c$ for fixed $\lambda=1/4$ and
$W=1$.  The line with the
arrow is the trajectory through parmeter space referred to in Fig. 5.
\bigskip

5)  The phase diagram as a function of temperature and position along
the representative trajectory through parameter space indicated
in Fig. 4, both in the
presence and absence of disorder.


\begin{references}
\bibitem{1dexample}  The one dimensional hard-core bose gas on the
lattice is easily mapped onto a spin 1/2 XXZ spin-chain problem, in
which the XY part of the Hamiltonian is ferromagnetic (reflecting
the nodelessness of a boson ground-state) with a magnitude $J_{xy}$
given by
the boson kinetic energy, and the Ising part $J_z$
reflects a nearest-
neighbor interaction between bosons, and can be ferromagnetic (attractive)
or antiferromagnetic (repulsive);  one can also include a chemical
potential, which translates into a magnetic field in the z direction.
At zero temperature and
zero magnetic field this has an XY (superconducting)
phase with power-law order for
$J_{xy} > |J_z|$, a period two Neel (charge-density wave) phase for
$J_z > J_{xy}$ and is ferromagnetic (phase separates) for $J_z < - J_{xy}$.
At finite temperature, of course, there are no ordered phases.
The quantum spherical model defined as in the present paper has
precisely these phases, in addition to a zero-temperature
quantum disordered phase.
\bibitem{alpha}  It is easy to see that so long as $\alpha_s$ and
$\alpha_p$ are both positive and non-zero, we can rescale the
spins and momenta, with a suitable rescaling of physical parameters
(including $\hbar$) to write the model in a form with $\alpha_s=
\alpha_p=1$.
\bibitem{plasma}
Note that for $\mu\ne 0$, $\omega_{\vec k} \sim \sqrt{\mu Q}/k$ as
$k \to 0$, which is unphysical;  this mode would be expected
to approach the plasma freqency as $k\to 0$ in the absence of
any addition degrees of freedom which might screen the Coulomb
interactions.  However, this unphysical aspect of the model
does not appear to have much effect on the central physical results.
For instance, if we imagine that the Coulomb interactions was
screened at long distances, this would remove this unphysical divergence
without affecting the physics of the model at finite $Q$;  effects
of $1/n$ corrections typically produce such screening and so might
be expected to remove this unphysical aspect of the model, as well.
In the superconducting state, this mode approaches a finite freqency,
$\omega_{\vec k} \to \sqrt{W Q}$ as
$k \to 0$, which has a natural interpretation as the plasma mode.
\end{references}
\end{document}